\documentclass[prd,twocolumn,floatfix,amsmath,nofootinbib,amssymb,floatfix,10pt]{revtex4-2}
\usepackage{graphicx,color,dcolumn,booktabs,bm}
\usepackage{txfonts}
\usepackage{amssymb}
\usepackage{ulem}
\usepackage{feynmf}   
\usepackage{slashed}  
\usepackage{cases}
\usepackage{color}
\usepackage{multirow}
\usepackage{threeparttable}
\usepackage{epstopdf}
\usepackage{enumerate}
\usepackage[colorlinks=true,
            citecolor=blue,
            anchorcolor=red,
            menucolor=red,
            linkcolor=red,
            filecolor=red,
            runcolor=red,
            urlcolor=blue,
            frenchlinks=red]{hyperref}

\begin{document}

\title{Finding Interaction Mechanism between Exotic Molecule and Conventional Hadron}

\author{Ri-Qing Qian$^{1,2,4}$}
\email{qianrq@lzu.edu.cn}
\author{Fu-Lai Wang$^{1,2,4}$}
\email{wangfl2016@lzu.edu.cn}
\author{Xiang Liu$^{1,2,3,4}$}
\email{xiangliu@lzu.edu.cn}
\affiliation{$^1$School of Physical Science and Technology, Lanzhou University, Lanzhou 730000, China\\
$^2$Lanzhou Center for Theoretical Physics,
Key Laboratory of Theoretical Physics of Gansu Province,
Key Laboratory of Quantum Theory and Applications of MoE,
Gansu Provincial Research Center for Basic Disciplines of Quantum Physics, Lanzhou University, Lanzhou 730000, China\\
$^3$MoE Frontiers Science Center for Rare Isotopes, Lanzhou University, Lanzhou 730000, China\\
$^4$Research Center for Hadron and CSR Physics, Lanzhou University and Institute of Modern Physics of CAS, Lanzhou 730000, China}

\begin{abstract}
An intriguing and important question arises: how do hadronic molecules interact with conventional hadrons? In this work, we propose a novel mechanism to address this issue, focusing on the interactions between hidden-charm molecular pentaquarks ($P_c$) and nucleons ($N$). By applying this mechanism, we derive the corresponding interaction potentials, enabling the prediction of the bound state properties. Our results indicate the possible existence of $P_cN$ bound states—an entirely new form of matter, ripe for exploration in the era of high-precision hadron spectroscopy. Notably, this mechanism extends beyond the $P_cN$ systems and can be applied to any hadronic molecules interacting with conventional hadrons, offering the potential for discovering a variety of novel bound states.
\end{abstract}
\maketitle

\section{Introduction}
The study of bound states {and related topics} is one of the central themes of modern physics. From molecules and atoms in atomic and molecular physics, to nuclei in nuclear physics, and hadrons in particle physics, the bound states are ubiquitous. At the scale of $10^{-15}$ meters, {the discovery of candidates of hadronic molecules—whether bound states, virtual states, or resonance systems arising from hadron-hadron interactions—has reignited the enthusiasm of the scientific community, making it a focal point in hadron physics.}

In recent years, the exploration of hadronic molecules has sparked extensive research into how conventional hadrons form exotic molecular states \cite{Hosaka:2016pey,Chen:2016qju,Guo:2017jvc,Liu:2019zoy,Brambilla:2019esw,Chen:2022asf}. Experiments such as those conducted by LHCb have reported several good candidates for hadronic molecular states, including the $P_c$ \cite{LHCb:2015yax,LHCb:2019kea}, $P_{cs}$ \cite{LHCb:2020jpq,LHCb:2022ogu}, and $T_{cc}$ \cite{LHCb:2021vvq,LHCb:2021auc} states. These discoveries, alongside other newly observed hadronic states, signal a new era in hadron spectroscopy, often referred to as the construction of ``Particle Zoo 2.0". The core of this endeavor is to deepen our understanding of the strong interaction, which exhibits inherently nonperturbative behavior. 

As we reach this new stage, an intriguing and important question arises: how do hadronic molecules interact with conventional hadrons? This open question remains to be explored further. 

In this work, we propose a novel mechanism to address this problem. Focusing on the interactions between hidden-charm molecular pentaquarks ($P_c$) and nucleons ($N$), we use this mechanism to derive the corresponding interaction potentials, which allows us to predict the bound state properties. Our results suggest the possible existence of the $P_cN$ bound states—a new form of matter worthy of exploration, particularly in the era of high-precision hadron spectroscopy. Importantly, this mechanism is not restricted to the $P_cN$ systems; it is applicable to all hadronic molecules interacting with conventional hadrons, potentially leading to the discovery of a variety of novel systems.

\section{Model for Interaction Dynamics}

\begin{figure}[htbp]
  \centering
  \includegraphics[width=8.6cm]{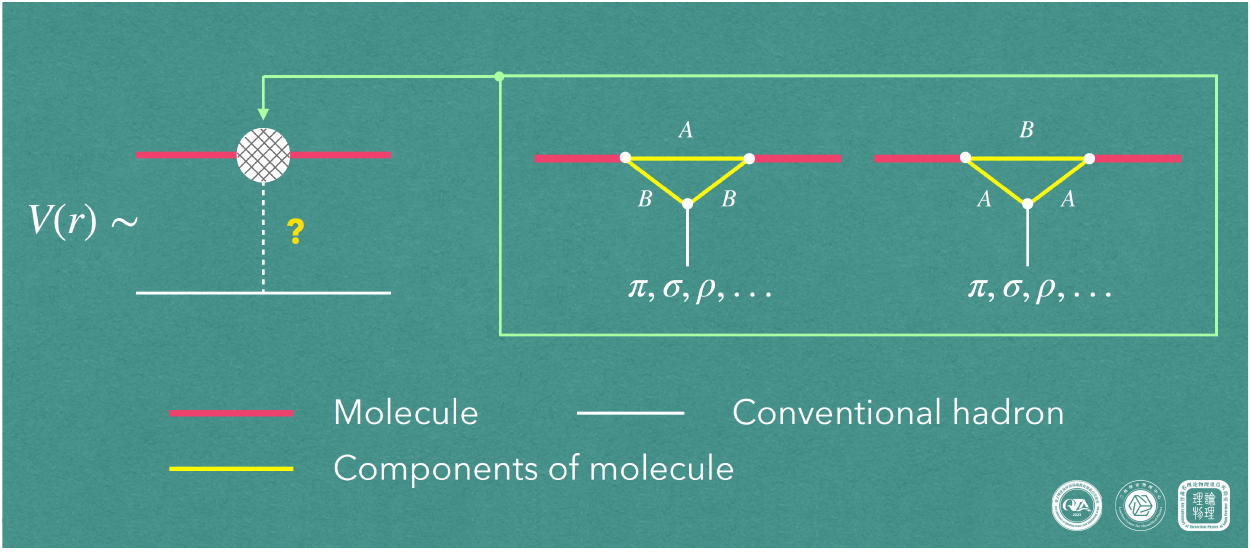}
  \caption{(Color online.) A novel and universal mechanism describing the interactions between the molecular state and the conventional hadron, and the interaction mechanisms in different scale.}\label{fig:Novelmechanism}
\end{figure}

The interaction mechanisms in different scale are quite different. From the photon exchange in the atomic scale, to the pion exchange theory of nucleon force, and further to the gluon exchange between the quark inside the hadron. 
In the hadronic scale, our understanding of the interactions mechanism between the conventional hadrons is promoted due to the discovery of various candidates of hadronic molecules \cite{Hosaka:2016pey,Chen:2016qju,Guo:2017jvc,Liu:2019zoy,Brambilla:2019esw,Chen:2022asf}, but the interactions between hadronic molecules and conventional hadrons remain to be explored further. Here, we propose a novel and universal mechanism describing the molecular state and conventional hadron interactions (see Fig.~\ref{fig:Novelmechanism}). It is well established that the molecular states are the loosely bound states \cite{Chen:2016qju}, which indicates that the constituent hadrons exist the large distance in the spatial distribution. Furthermore, there is the strong coupling between the constituent hadrons and the molecular state \cite{Guo:2017jvc}. Consequently, the interactions between the molecular states and the conventional hadrons can occur through the constituent hadron and conventional hadron interactions. The one-boson-exchange (OBE) mechanism is the successful approach to study the interactions between the conventional hadrons \cite{Chen:2016qju}. Thus, the constituents of the molecular states can interact with the conventional hadrons via the exchange of the bosons, and the coupling between the molecular states and the exchanged bosons can be calculated via a loop diagram.

\begin{figure}[htbp]
  \centering
  \includegraphics[width=8 cm]{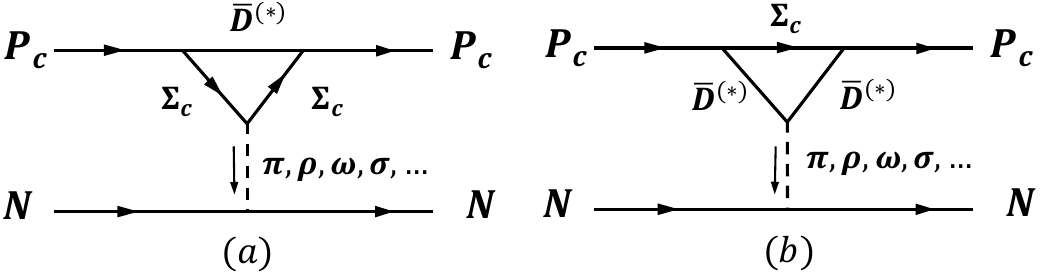}
  \caption{The mechanism that depict the $P_cN$ interactions. Here, the $P_c$ state is treated as the $\Sigma_c\bar{D}^{(*)}$ molecular state.}\label{fig:Schematic}
\end{figure}

For the experimental discoveries of new hadron states, the observed $P_c(4312)$, $P_c(4440)$, and $P_c(4457)$ states \cite{LHCb:2019kea} provides the compelling experimental evidence for the existence of the $\Sigma_c \bar D^{(*)}$ molecular pentaquarks \cite{Chen:2015loa,Chen:2019asm,Chen:2019bip,Liu:2019tjn,Xiao:2019aya,Meng:2019ilv,Du:2019pij}. Thus, the $P_c$ states can interact with nucleon via the $\Sigma_cN$ and $\bar{D}^{(*)}N$ interactions. Within the OBE mechanism, it is well established that the nucleon can interact with $\Sigma_c$ and $\bar{D}^{(*)}$ via the exchange of the light mesons (See Fig.~\ref{fig:Schematic}). Thus, we concentrate on the interactions between hidden-charm molecular pentaquarks and nucleons in the following.

{It is more appropriate to classify the $P_cN$ system as a three-body system, given that the observed $P_c$ states are interpreted as $\Sigma_c\bar{D}^{(*)}$ molecular states. Several methods exist for solving the three-body bound state problem. The interaction between clusters can be studied using the Resonating Group Method~\cite{Entem:2000mq,Fernandez:1993hx,RGMbook}, while the three-body system can also be solved directly using various method such as the Gaussian Expansion Method (GEM) \cite{Hiyama:2003cu,Wu:2021kbu,Zhu:2024hgm}, variational approach~\cite{Bayar:2012dd} and Born-Oppenheimer approximation~\cite{Ma:2018vhp,Ma:2017ery}. Additionally, the standard framework for addressing three-body scattering problems is the Faddeev equation \cite{Faddeev:1960su}, and the approach we adopt here can be related to the Faddeev formalism.} The Faddeev equation decomposes the total three-body $T$-matrix into three components:
\begin{eqnarray}
  T=T^1+T^2+T^3 \;.
\end{eqnarray}
The components $T^i$ satisfy a set of coupled equations:
\begin{eqnarray}
  T^i = t^i + t^ig^{ij}T^J + t^ig^{ik}T^k \;,
\end{eqnarray}
where $t^i$ represents the two-body $T$-matrix for interactions between the pair $(jk)$, and 
$g^{ij}$ denotes the three-body propagator or Green function\footnote{{It should be noted that the interaction considered here is pairwise, but not genuine three-body interaction that characterize the three-body force~\cite{Hammer:2012id}.}}.
In many practical scenarios, solving the full set of Faddeev equations is computationally prohibitive. However, when two of the three particles exhibit a significantly stronger mutual correlation compared to their interactions with the third particle—resulting in the formation of a bound state—the Fixed Center Approximation (FCA) provides a viable simplification \cite{Barrett:1999cw,Deloff:1999gc,Kamalov:2000iy,MartinezTorres:2020hus}. Within the FCA, the Faddeev equations reduce to
\begin{eqnarray}
  T^1 &=& t^1 + t^1G_0T^2 ,  \nonumber\\
  T^2 &=& t^2 + t^2G_0T^1 , \\
  T &=& T^1+T^2 \;. \nonumber
\end{eqnarray}
Here, $T^i$ represents the two-body scattering amplitudes involving the third particle and the bound pair. For this study, we neglect higher-order contributions such as double scattering terms 
$t^iG_0t^j$, focusing instead on the single-scattering processes $t^i$, which are further approximated using the one-boson-exchange amplitude. The resulting three-body scattering amplitudes correspond directly to the diagrams shown in Fig.~\ref{fig:Schematic}.
  
{From the above discussions, we find that three-body system are inherently complex, requiring a detailed understanding of the underlying two-particle interactions. Prior to discussing the $P_cN$ interactions, it is pertinent to undertake a brief comparison of the $\bar{D}^{(*)}N$ and $\Sigma_cN$ interactions within the OBE model with existing theoretical works.

The $\bar{D}N$ interaction has been extensively studied using various theoretical approaches \cite{Hofmann:2005sw,Gamermann:2010zz,Haidenbauer:2007jq,Yasui:2009bz,Yamaguchi:2011xb,Yamaguchi:2011qw,Carames:2012bd,Fontoura:2012mz}. For $J^P=(1/2)^-$ $\bar{D}N$ system, the $\bar{D}N$ interaction is either insufficient to form the loosely bound state~\cite{Hofmann:2005sw,Haidenbauer:2007jq,Carames:2012bd,Fontoura:2012mz} or only capable of forming very shallow bound states~\cite{Gamermann:2010zz,Yamaguchi:2011xb} for $J^P=1/2^-(I=0)$ $\bar{D}N$ system.
The OBE potential for the $\bar{D}N$ system can be obtained from the $DN$ potential \cite{Luo:2022cun} by the $G$-parity transformation, this potential does not yield the loosely bound state solutions in both the $I=0$ and $I=1$ channels for the $\bar{D}N$ system. Although the potential for the $I=0$ $\bar{D}N$ channel is attractive, it is insufficient to form a loosely bound state. This conclusion is in qualitative alignment with the findings of previous studies. For the $\Sigma_cN$ interaction, it is notable that the predictions for the $\Sigma_cN$ interaction vary across theoretical models. The OBE approach~\cite{Liu:2011xc} predicts the existence of the bound states for both $I(J^P)=1/2(0^+)$ and $I(J^P)=1/2(1^+)$ channels with reasonable cutoff values. These findings are consistent with the quark delocalization color-screening model~\cite{Huang:2013zva}. In contrast, studies using chiral effective field theory~\cite{Haidenbauer:2017dua} and constituent quark model~\cite{Garcilazo:2019ryw} did not yield any evidence of the bound states. In light of the disparate predictions, it is challenging to determine the most appropriate method to describe the $\Sigma_cN$ interaction. In order to ensure consistency, we adopt the OBE framework to analyze the $\bar{D}N$, $\Sigma_cN$, and $P_cN$ interactions.}

Now, we discuss the interactions between hidden-charm molecular pentaquarks and nucleons. Here, we employ the effective Lagrangian approach to describe the relevant schematic diagrams. Among them, the $P_c\Sigma_c\bar{D}^{(*)}$ interactions can be described by the following effective Lagrangians \cite{Lin:2019qiv}
\begin{equation}
\begin{split}
   \mathcal{L}_{\bar{D}\Sigma_cP_c^{1/2^-}} & = g_{\bar{D}\Sigma_cP_c^{1/2^-}}\bar{P}_c\Sigma_c\bar{D}\,, \\
   \mathcal{L}_{\bar{D}^*\Sigma_cP_c^{1/2^-}} & = g_{\bar{D}^*\Sigma_cP_c^{1/2^-}}\bar{P}_c\gamma^5\tilde{\gamma}^\mu \Sigma_c\bar{D}^*_\mu \,, \\
   \mathcal{L}_{\bar{D}^*\Sigma_cP_c^{3/2^-}} & = g_{\bar{D}^*\Sigma_cP_c^{3/2^-}}\bar{P}_c^{\mu}\Sigma_c\bar{D}^*_\mu \,,
\end{split}
\end{equation}
where $\tilde{\gamma}^\mu\equiv (g^{\mu\nu}-p^\mu p^\nu/p^2)\gamma_\nu$. These Lagrangians can be expanded by the flavor wave functions of the isospin-1/2 $\Sigma_c\bar{D}^{(*)}$ systems, i.e.,
\begin{equation}\label{eq:wavefunction}
  \begin{split}
 \left|\frac{1}{2},+\frac{1}{2}\right\rangle &= \sqrt{\frac{2}{3}}\left|\Sigma_c^{++}D^{(*)-}\right\rangle - \sqrt{\frac{1}{3}}\left|\Sigma_c^+\bar{D}^{(*)0}\right\rangle\,, \\
\left|\frac{1}{2},-\frac{1}{2}\right\rangle &= \sqrt{\frac{1}{3}}\left|\Sigma_c^+D^{{(*)}-}\right\rangle - \sqrt{\frac{2}{3}}\left|\Sigma_c^0\bar{D}^{(*)0}\right\rangle\,.
  \end{split}
\end{equation}
{In the limit $\sqrt{2\mu E_B}\ll m_\pi$, the coupling constants of molecule with its components can be determined by its binding energy~\cite{Weinberg:1962hj}. We determine the coupling $g_{\bar{D}^{(*)}\Sigma_c P_c^{1/2(3/2)^-}}$ in this way, though the binding momentum $E_B\sim 10$ MeV of $P_c$ states implying that the condition $\sqrt{2\mu E_B}\ll m_\pi$ is not strictly satisfied and this approximation may introduce uncertainties. In this way, the coupling constants have the form \cite{Lin:2019qiv}}
\begin{equation}
  g_{\bar{D}^{(*)}\Sigma_cP_c^{1/2(3/2)^-}} = \sqrt{\frac{8\sqrt{2|E_B|}m_1m_2\pi}{(m_1m_2/(m_1+m_2))^{3/2}}}\sqrt{\frac{1}{MF}} \,.
\end{equation}
Here, $m_1$ and $m_2$ denote the masses of $\Sigma_c$ and $\bar{D}^{(*)}$, respectively. $M$ are $2m_1$, $6m_1$, and $4/3m_1$ for the spin-1/2 $\Sigma_c\bar{D}$ molecule, the spin-1/2 $\Sigma_c\bar{D}^*$ molecule, and the spin-3/2 $\Sigma_c\bar{D}^*$ molecule, respectively. $F$ are 1 and 3/2 for the spin-1/2 and spin-3/2 molecules, respectively.

In addition, the effective Lagrangians describing the interactions between the hadrons $\mathcal{B}_6/\bar{D}^{(*)}/N$ and the light mesons $M$ are constructed as  \cite{Wise:1992hn,Casalbuoni:1992gi,Yan:1992gz,Casalbuoni:1996pg,Bando:1987br,Harada:2003jx,Wang:2023iox,Wang:2021hql}
\begin{eqnarray}
  \mathcal{L}_{\bar{\mathcal{B}}_6\mathcal{B}_6 M} &= & i\frac{g_1}{2f_\pi}\epsilon^{\mu\nu\lambda\kappa}v_\kappa \langle \bar{\mathcal{B}}_6\gamma_\mu\gamma_\lambda\partial_\nu P\mathcal{B}_6\rangle\nonumber\\
  &&-l_S \langle \bar{\mathcal{B}}_6\sigma\mathcal{B}_6\rangle-\frac{\beta_Sf_V}{\sqrt{2}} \langle \bar{\mathcal{B}}_6v\cdot V\mathcal{B}_6\rangle\nonumber\\
  &&-i\frac{\lambda_S f_V}{3\sqrt{2}}\langle\bar{\mathcal{B}}_6\gamma_{\mu}\gamma_{\nu} \left(\partial^{\mu}\mathbb{V}^{\nu}-\partial^{\nu}\mathbb{V}^{\mu}\right)
    \mathcal{B}_6\rangle,\\
  \mathcal{L}_{\bar{\mathcal{D}}^{(*)}\bar{\mathcal{D}}^{(*)}M} &=& \frac{2ig}{f_\pi}v^\alpha\epsilon_{\alpha\mu\nu\lambda} \bar{\mathcal{D}}_a^{*\mu\dagger}\bar{\mathcal{D}}_b^{*\lambda}\partial^\nu P_{ab} \nonumber\\
  & &+\frac{2g}{f_\pi}(\bar{\mathcal{D}}_a^{*\mu\dagger}\bar{\mathcal{D}}_b + \bar{\mathcal{D}}_a^\dagger\bar{\mathcal{D}}_b^{*\mu})\partial_\mu P_{ab}\nonumber\\
  &&-2g_S\bar{\mathcal{D}}_a\bar{\mathcal{D}}_a^\dagger \sigma + 2g_S \bar{\mathcal{D}}^*_{a\mu}\bar{\mathcal{D}}_a^{*\mu\dagger}\sigma\nonumber\\
&& +\sqrt{2}\beta f_V \bar{\mathcal{D}}_a \bar{\mathcal{D}}_b^\dagger v\cdot V_{ab}-\sqrt{2}\beta f_V \bar{\mathcal{D}}^*_{a\mu}\bar{\mathcal{D}}_b^{*\mu\dagger}v\cdot V_{ab} \nonumber\\
  & &-2\sqrt{2}i\lambda f_V \bar{\mathcal{D}}_a^{*\mu\dagger}\bar{\mathcal{D}}_b^{*\nu} (\partial_\mu V_\nu-\partial_\nu V_\mu)_{ab} \nonumber\\
  & &-2\sqrt{2}\lambda f_V v^\lambda \epsilon_{\lambda\mu\alpha\beta}(\bar{\mathcal{D}}_a^{*\mu\dagger} \bar{\mathcal{D}}_b + \bar{\mathcal{D}}_a^{\dagger}\bar{\mathcal{D}}_b^{*\mu})\partial^\alpha V_{ab}^\beta \,,\\
  \mathcal{L}_{NNM} &= &\; g_{\sigma NN} \bar{N}\sigma N + \sqrt{2}g_{\pi NN}\bar{N}i\gamma_5PN \nonumber\\
  &&+ \sqrt{2}g_{\rho NN}\bar{N}\gamma_\mu V^\mu N + \frac{f_{\rho NN}}{\sqrt{2}m_N}\bar{N}\sigma_{\mu\nu}\partial^\mu V^\nu N,
\end{eqnarray}
where the matrices for $\mathcal{B}_6$, $P$, and $V_\mu$ are defined as
\begin{equation}
\begin{split}
  \mathcal{B}_6 &= \left(
                          \begin{array}{cc}
                            \Sigma_c^{++} & \frac{\Sigma_c^{+}}{\sqrt{2}} \\
                            \frac{\Sigma_c^{+}}{\sqrt{2}} & \Sigma_c^{0}\\
                          \end{array}
                        \right) \,,\quad\quad
  P = \left(
                  \begin{array}{cc}
                    \frac{\pi^0}{\sqrt{2}} + \frac{\eta}{\sqrt{6}} & \pi^+  \\
                    \pi^-                                          & -\frac{\pi^0}{\sqrt{2}} + \frac{\eta}{\sqrt{6}} \\
                  \end{array}
                \right) \, ,\quad \\
 V_\mu &= \left(
                      \begin{array}{cc}
                        \frac{\rho^0}{\sqrt{2}}+\frac{\omega}{\sqrt{2}}  &\rho^+\\
                        \rho^- & -\frac{\rho^0}{\sqrt{2}}+\frac{\omega}{\sqrt{2}}  \\
                      \end{array}
                    \right)_\mu \,,
\end{split}
\end{equation}
respectively. {The normalization for the above charmed baryon field $\mathcal{B}$ is $\langle 0|\mathcal{B}|cqq\rangle = \sqrt{2m_{\mathcal{B}}}(\chi,\frac{\sigma\cdot\boldsymbol{p}}{2m_{\mathcal{B}}}\chi)^T$, and the normalization for the $\mathcal{D}$ and $\mathcal{D}^*$ mesons are $\langle 0|\mathcal{D}|c\bar{q}(0^-)\rangle = \sqrt{m_D}$ and $\langle 0|\mathcal{D}^{*\mu}|c\bar{q}(1^-)\rangle = \sqrt{m_{D^*}}\epsilon^\mu$, respectively.}
The coupling constants in the above effective Lagrangians are the fundamental inputs, and we take $l_S = 6.20$, $g_1=0.94$, $\beta_Sf_V = 12.00$, $\lambda_Sf_V = 19.20\; \rm{GeV}^{-1}$  \cite{Wang:2023iox}, $g_S = 0.76$, $g=0.59$, $\beta f_V = -5.25$, $\lambda f_V = -3.27\,\rm{GeV}^{-1}$ \cite{Wang:2021hql}, $\frac{g^2_{\sigma NN}}{4\pi} = 5.69$, $\frac{g^2_{\pi NN}}{4\pi} = 13.60$,  $\frac{g^2_{\rho NN}}{4\pi} = 0.84$, and $\frac{f_{\rho NN}}{g_{\rho NN}} = 6.10$ \cite{Machleidt:2000ge,Machleidt:1987hj,Cao:2010km}, where the phase factors between the relevant coupling constants can be determined by the quark model \cite{Riska:2000gd}.


Based on Fig.~\ref{fig:Schematic}, the scattering amplitude of the $P_cN\to P_cN$ process can be expressed as 
\begin{eqnarray}
i\mathcal{M}(q)=\sum_{M=\sigma,\,P,\,V}\mathcal{V}^{P_cP_cM}_{(\mu)} P_{M}^{(\mu\nu)} \mathcal{V}^{NNM}_{(\nu)}.
\end{eqnarray}
Here, $\mathcal{V}^{P_cP_cM}_{(\mu)}$ and $\mathcal{V}^{NNM}_{(\nu)}$ denote the $P_c$-$P_c$-$M$ and $N$-$N$-$M$ coupling vertices, respectively. $P_{M}^{(\mu\nu)}$ is the propagator of the exchanged light meson $M$. For $\mathcal{V}^{P_cP_cM}_{(\mu)}$, there contains the $\Sigma_c\bar{D}^{(*)}$ triangle loop, which can be calculated by performing the loop integral. 

In the following, we illustrate how to deduce the $P_c^+(4312)P_c^+(4312)\pi^0$ vertex explicitly, which can be written as
\begin{eqnarray}
     & &\mathcal{V}^{P_c^+(4312)P_c^+(4312)\pi^0}\nonumber\\
     &&= i^3 \frac{1}{\sqrt{2}} g_{\bar{D}\Sigma_cP_c^{1/2^-}}^2\int\frac{d^4q_3}{(2\pi)^4}\,\bar{u}(p_1)\left(\slashed{q}_1+m_{\Sigma_c}\right)i\frac{g_1}{2f_\pi} \epsilon^{\mu\nu\lambda\kappa}v_\kappa\gamma_\mu\gamma_{\lambda}iq_{\nu}\nonumber\\
     & &\quad \times \left( \slashed{q}_2+m_{\Sigma_c} \right)u(p_2)\frac{i}{q_1^2-m_{\Sigma_c}^2}\frac{i}{q_2^2-m_{\Sigma_c}^2}\frac{i}{q_3^2-m_D^2}\mathcal{F}_{l}^2(q_3^2,m_D^2)\nonumber\\
      &&\equiv \frac{1}{\sqrt{2}}\mathcal{V}_{P},
\end{eqnarray}
where $\mathcal{F}_{l}(q_3^2,\,m_D^2)=(\tilde{\Lambda}^2-m_D^2)^2/(\tilde{\Lambda}^2-q_3^2)^2$ is the form factor introduced to consider the off-shell effect of the constituent hadrons of the molecule state and suppress the divergence of the loop integral. Usually, $\tilde{\Lambda}$ is parameterized as $\tilde{\Lambda}=m_D+\alpha \Lambda_{\rm QCD}$ with $\Lambda_{\rm QCD}=0.22\,{\rm GeV}$, and one expects that the parameter $\alpha$ should be around 1 \cite{Cheng:2004ru}. Thus, $\tilde{\Lambda}$ is not significantly different from the mass of the $D$ and the off-shell part of the $D$ contributes less to the loop integral. Similarly, we can get the following relations
\begin{eqnarray}
    \mathcal{V}^{P_c^0(4312)P_c^0(4312)\pi^0} &=& -\mathcal{V}^{P_c^+(4312)P_c^+(4312)\pi^0}= -\frac{1}{\sqrt{2}}\mathcal{V}_{P}\,,\\
    \mathcal{V}^{P_c^+(4312)P_c^0(4312)\pi^-}&=& \mathcal{V}^{P_c^0(4312)P_c^+(4312)\pi^+}=\mathcal{V}_{P}\,, \\
    \mathcal{V}^{P_c^0(4312)P_c^0(4312)\eta} &=& \mathcal{V}^{P_c^+(4312)P_c^+(4312)\eta}=\frac{\sqrt{6}}{4} \mathcal{V}_{P}\,,\\
    \mathcal{V}^{P_c^0(4312)P_c^0(4312)\rho^0} &=& -\mathcal{V}^{P_c^+(4312)P_c^+(4312)\rho^0} \nonumber\\
    &=& -\frac{1}{\sqrt{2}}\left(\mathcal{V}_{{V 1}}^{\mu} -\mathcal{V}_{{V 2}}^{\mu}\right)\,, \\
    \mathcal{V}^{P_c^+(4312)P_c^0(4312)\rho^-}&=&\mathcal{V}^{P_c^0(4312)P_c^+(4312)\rho^+} \nonumber\\
    &=&\mathcal{V}_{V 1}^{\mu} - \frac{1}{2}\mathcal{V}_{V 2}^{\mu}\,,\\
    \mathcal{V}^{P_c^+(4312)P_c^+(4312)\omega}&=& \mathcal{V}^{P_c^0(4312)P_c^0(4312)\omega} \nonumber\\
    &=&\frac{3\sqrt{2}}{4}(\mathcal{V}_{V1}^{\mu} + \mathcal{V}_{V 2}^{\mu})\,,\\
    \mathcal{V}^{P_c^0(4312)P_c^0(4312)\sigma} &=&\mathcal{V}^{P_c^+(4312)P_c^+(4312)\sigma}\nonumber \\
    &=&\frac{3}{2}\left(\mathcal{V}_{S 1} +\mathcal{V}_{S 2}\right),
\end{eqnarray}
where
\begin{eqnarray}
 \mathcal{V}_{V 1}^{\mu}&=& i^3 g_{\bar{D}\Sigma_cP_c^{1/2^-}}^2\int \frac{d^4q_3}{(2\pi)^4} \bar{u}(p_1) \left(\slashed{q}_1+m_{\Sigma_c}\right)
  \left[-\frac{\beta_Sf_V}{\sqrt{2}}v^\mu\right. \nonumber\\
       &&+\left.\frac{\lambda_Sf_V}{3\sqrt{2}}\gamma_\alpha\gamma_\beta(q^\alpha g^{\mu\beta}-q^\beta g^{\mu\alpha})\right] \left(\slashed{q}_2+m_{\Sigma_c}\right) u(p_2)\nonumber\\
  && \times \frac{i}{q_1^2-m_{\Sigma_c}^2} \frac{i}{q_2^2-m_{\Sigma_c}^2} \frac{i}{q_3^2-m_D^2} \mathcal{F}_l^2(q_3^2,m_D^2)\,,\\
  \mathcal{V}_{V 2}^{\mu}& =& i^3g_{\bar{D}\Sigma_cP_c^{1/2^-}}^2\int\frac{d^4q_3}{(2\pi)^4}\bar{u}(p_1)\left(\slashed{q}_3+m_{\Sigma_c}\right)u(p_2) \nonumber\\
  && \times\sqrt{2}\beta m_D f_V v^\mu\frac{i}{q_1^2-m_D^2}\frac{i}{q_2^2-m_D^2}\frac{i}{q_3^2-m_{\Sigma_c}^2}\mathcal{F}_l^2(q_3^2,m_{\Sigma_c}^2)\,,\nonumber\\\\
\mathcal{V}_{S 1} &= &i^3 g_{\bar{D}\Sigma_cP_c^{1/2^-}}^2(-l_S)\int \frac{d^4q_3}{(2\pi)^4} \bar{u}(p_1)\left(\slashed{q}_1+m_{\Sigma_c}\right) \nonumber\\
  && \times\left(\slashed{q}_2+m_{\Sigma_c}\right)u(p_2)\frac{i}{q_1^2-m_{\Sigma_c}^2} \frac{i}{q_2^2-m^2_{\Sigma_c}}\frac{i}{q_3^2-m_D^2} \nonumber\\
  && \times\, \mathcal{F}_l^2(q_3^2,m_D^2)\,,\\
\mathcal{V}_{S 2} &= &  i^3 g_{\bar{D}\Sigma_cP_c^{1/2^-}}^2(-2m_D g_S)\int \frac{d^4q_3}{(2\pi)^4}\bar{u}(p_1)\left(\slashed{q}_3+m_{\Sigma_c}\right)u(p_2)\nonumber\\
  && \times\frac{i}{q_1^2-m_D^2}\frac{i}{q_2^2-m_D^2}\frac{i}{q_3^2-m_{\Sigma_c}^2} \mathcal{F}_l^2(q_3^2,m_{\Sigma_c}^2)\,.
\end{eqnarray}
After performing the loop integral, the explicit expressions of the coupling vertices can be obtained. In the case of the $P_c(4312)P_c(4312)P$ vertex,  we have
\begin{equation}\label{eq:vertex_4312_P}
  \mathcal{V}_{P} = i \frac{g_{_P}(q^2)}{m_{P_c}}\,\bar{u}(p_1)\gamma^\lambda\gamma^\mu u(p_2)\epsilon_{\lambda\mu\alpha\beta}q^{\alpha}v^\beta\,,
\end{equation}
where $g_{_P}(q^2)$ is the coefficient as the function of the four-momentum square $q^2$ of the exchange pseudoscalar mesons. In addition, we have
\begin{eqnarray}
  \mathcal{V}_{V 1}^{\mu} & =& i g_{_{V1}}(q^2) \bar{u}(p_1) u(p_2) v^\mu+ig_{_{V3}}(q^2)\bar{u}(p_1)\gamma^\mu u(p_2) \nonumber\\
                               & &+\, i\frac{g_{_{V2}}(q^2)}{m_{P_c}} \bar{u}(p_1)u(p_2)(p_1^\mu+p_2^\mu)\,,\\
  \mathcal{V}_{V 2}^{\mu} & = &ig_{_{V4}}(q^2) \bar{u}(p_1) u(p_2) v^\mu \,,\\
  \mathcal{V}_{S 1} & =& i g_{_{S 1}}(q^2) \bar{u}(p_1)u(p_2)\,,\\
  \mathcal{V}_{S 2} & =& i g_{_{S 2}}(q^2) \bar{u}(p_1)u(p_2)\,.\label{eq:vertex_4312_S}
\end{eqnarray}

\begin{figure}[htbp]
  \centering
  \includegraphics[width=8.6cm]{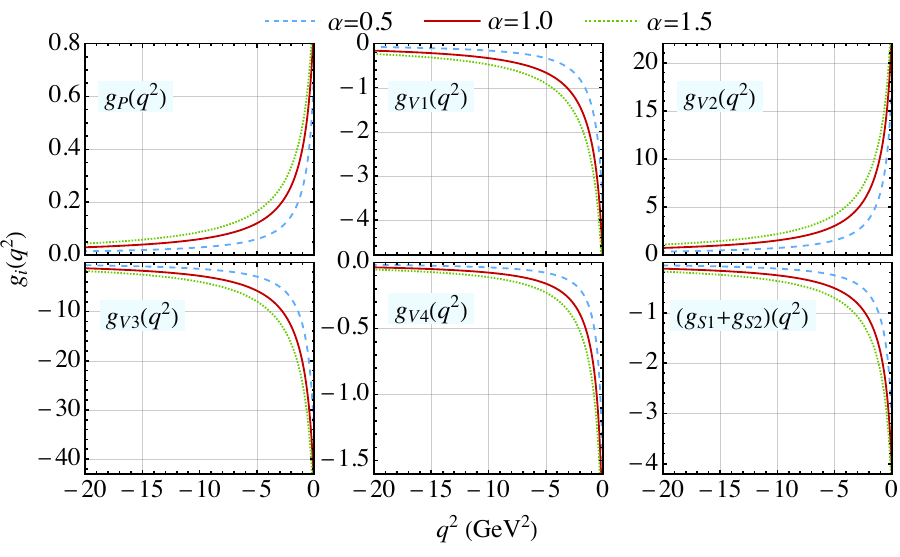}
  \caption{$\alpha$ dependence of various coupling coefficients $g_i(q^2)$ for the $P_c(4312)P_c(4312)M$ vertexes.}\label{eq:4312vertexes}
\end{figure}

In Fig.~\ref{eq:4312vertexes}, we show $\alpha$ dependence of various coupling coefficients listed in Eqs.~\eqref{eq:vertex_4312_P}-\eqref{eq:vertex_4312_S}. There are 1 coupling coefficient for the pseudoscalar meson coupling, 4 coupling coefficients for the vector meson coupling, and 1 scalar meson coupling coefficient, while the coupling coefficients show a regular dependence on $q^2$, and $q^2<0$ as a usual $t$-channel exchange process. As shown in Fig.~\ref{eq:4312vertexes}, the dependence of various coupling coefficients $g_{_{P,Vi,Si}}$ on the parameter $\alpha$ is not significant. Thus, we take $\alpha$ to be the typical value of 1 in the subsequent analysis.

The scattering amplitude of the $P_cN\to P_cN$ process in terms of the isospin basis can be obtained from the aforementioned vertex structures. For the $P_cN$ systems, there exist two isospin states, i.e.,
\begin{equation}
  \begin{split}
     |0,0\rangle = \frac{1}{\sqrt{2}}\left(P_c^+n - P_c^0p\right) ~~{\rm and}~~~
     |1,1\rangle = P_c^+p\,,
  \end{split}
\end{equation}
and the isospin amplitudes $\langle I|T|I\rangle$ for the $P_cN$ systems can be decomposed as
\begin{eqnarray}
   \langle 0|T|0\rangle &= & \frac{1}{2}\langle P_c^+n|T|P_c^+n\rangle + \frac{1}{2}\langle P_c^0p|T|P_c^0p\rangle\nonumber\\
   &&-\frac{1}{2}\langle P_c^0p|T|P_c^+n\rangle - \frac{1}{2} \langle P_c^+n|T|P_c^0p\rangle\,, \\
  \langle 1|T|1\rangle&=& \langle P_c^+p|T|P_c^+p\rangle\,.
\end{eqnarray}
Utilizing the aforementioned discussions, the scattering amplitude of the $P_c(4312) N\to P_c(4312) N$ process by exchanging the pseudoscalar meson $P$ can be expressed as
\begin{eqnarray}\label{eq:amp_pi}
    i\mathcal{M}_{P}(\boldsymbol{q})= \; C_{{P},\,I}\mathcal{V}_P\frac{i}{q^2-m_{P}^2}  i\sqrt{2}g_{NN\pi}\bar{u}(p_3)i\gamma_5 u(p_4)\,,
\end{eqnarray}
where $C_{\pi,0}=-{3}/{2}$, $C_{\pi,1}={1}/{2}$, and $C_{\eta}={1}/{4}$. For the vector meson exchange, we can get
\begin{eqnarray}
     i\mathcal{M}_{\mathcal{\rho}}(\boldsymbol{q})
   &= &\; C_{\mathcal{\rho},\,I}\left(\mathcal{V}_{V 1}^{\mu} - \frac{1}{2}\mathcal{V}_{V 2}^{\mu} \right) \frac{-ig_{\mu\nu}}{q^2-m_{\mathcal{V}}^2}i\bar{u}(p_3) \nonumber\\
     &&\times \left[ \sqrt{2}g_{\rho NN}\gamma^\nu + \frac{f_{\rho NN}}{\sqrt{2}m_N}\sigma^{\nu\lambda}(-iq_\lambda) \right]u(p_4) \,,\\
     i\mathcal{M}_{\mathcal{\omega}}(\boldsymbol{q})
   &= &\;C_{\mathcal{\omega}}\left(\mathcal{V}_{V 1}^{\mu} + \mathcal{V}_{V 2}^{\mu} \right) \frac{-ig_{\mu\nu}}{q^2-m_{\mathcal{V}}^2}i\bar{u}(p_3) \nonumber\\
     &&\times \left[ \sqrt{2}g_{\rho NN}\gamma^\nu + \frac{f_{\rho NN}}{\sqrt{2}m_N}\sigma^{\nu\lambda}(-iq_\lambda) \right]u(p_4) \,,
\end{eqnarray}
where $C_{\rho,0}=-{3}/{2}$, $C_{\rho,1}={1}/{2}$, and $C_{\omega}={3}/{4}$. For the scalar meson exchange, we have
\begin{eqnarray}\label{eq:amp_sigma}
    i\mathcal{M}_\sigma(\boldsymbol{q})
  =  \frac{3}{2}\left( \mathcal{V}_{S 1}+\mathcal{V}_{S 2} \right) \frac{i}{q^2-m_\sigma^2} ig_{\sigma NN}\bar{u}(p_3) u(p_4)\,.
\end{eqnarray}
Taking the $\pi$ exchange potential of the $P_c(4312)N$ system as example, in the non-relativistic limit, the scattering amplitude in the momentum space can be expressed as
\begin{eqnarray}
i\mathcal{M}_{{\pi}}(\boldsymbol{q})
     &\approx & 4C_{\pi,\,I} g_{P}(q^2)\chi_1^\dagger \boldsymbol{\sigma}\cdot\boldsymbol{q}\chi_2 \frac{i}{q^2-m_\pi^2}\sqrt{2}g_{NN\pi}\chi_3^\dagger\boldsymbol{\sigma}\cdot\boldsymbol{q}\chi_4 \nonumber\\
     &\approx& -4\sqrt{2}C_{\pi,\,I}i g_{P}(-\boldsymbol{ q}^2)g_{NN\pi} \frac{(\boldsymbol{\sigma}_1\cdot\boldsymbol{q})(\boldsymbol{\sigma}_2\cdot\boldsymbol{q})}{\boldsymbol{q}^2+m_\pi^2}\,.
\end{eqnarray}
{The other related potentials in the momentum space for the $P_c(4312)N$ system after non-relativistic reduction are presented in the following:
\begin{eqnarray}
 && ig(q^2)\bar{u}(p_1)u(p_2)v^\mu \frac{-ig_{\mu\nu}}{q^2-m_\rho^2} i\bar{u}(p_3)\sqrt{2}g_{\rho NN}\gamma^\mu u(p_4) \nonumber\\
     &&\approx  -i4\sqrt{2}m_{P_c}m_N g(-\boldsymbol{q}^2)g_{\rho NN} \frac{1}{\boldsymbol{q}^2+m_\rho^2}\,,\\
       && ig(q^2)\bar{u}(p_1)u(p_2)v^\mu \frac{-ig_{\mu\nu}}{q^2-m_\rho^2} i\bar{u}(p_3)\frac{f_{\rho NN}}{\sqrt{2}m_N}\sigma^{\nu\lambda}(-iq_\lambda) u(p_4) \nonumber\\
      &&\approx -(2m_{P_c})g(-\boldsymbol{q}^2)\frac{f_{\rho NN}}{\sqrt{2}m_N} \frac{1}{\boldsymbol{q}^2+m_\rho^2}\left[ 2(\boldsymbol{\sigma}_2\times\boldsymbol{k})\cdot\boldsymbol{q} + i \boldsymbol{q}^2 \right]\,,\\
    && i\frac{g(q^2)}{m_{P_c}}\bar{u}(p_1)u(p_2)(p_1^\mu + p_2^\mu) \frac{-i g_{\mu\nu}}{q^2-m_\rho^2} i\bar{u}(p_3)\sqrt{2}g_{\rho NN}\gamma^\nu u(p_4) \nonumber\\
    &&\approx -8\sqrt{2}i\,m_{P_c}m_N g(-\boldsymbol{q}^2)g_{\rho NN} \frac{1}{\boldsymbol{q}^2+m_\rho^2} \nonumber\\
    &&\quad+4\sqrt{2}ig(-\boldsymbol{q}^2)g_{\rho NN} \frac{1}{\boldsymbol{q}^2+m_\rho^2} \left[ -2\boldsymbol{k}^2 + i(\boldsymbol{\sigma}_2\times\boldsymbol{q})\cdot\boldsymbol{k} \right],\\
    && i\frac{g(q^2)}{m_{P_c}} \bar{u}(p_1)u(p_2)(p_1^\mu + p_2^\mu) \frac{-i g_{\mu\nu}}{q^2-m_\rho^2} i\bar{u}(p_3)\frac{f_{\rho NN}}{\sqrt{2}m_N} \sigma^{\nu\lambda}(-iq_\lambda) u(p_4) \nonumber\\
    &&\approx -4g(-\boldsymbol{q}^2) m_{P_c} \frac{f_{\rho NN}}{\sqrt{2}m_N} \frac{1}{\boldsymbol{q}^2+m_\rho^2} \left[ 2(\boldsymbol{\sigma}_2\times \boldsymbol{k})\cdot\boldsymbol{q} + i\boldsymbol{q}^2 \right] \nonumber\\
    &&\quad+8g(-\boldsymbol{q}^2) \frac{f_{\rho NN}}{\sqrt{2}} \frac{1}{\boldsymbol{q}^2+m_N^2} (\boldsymbol{\sigma}_2\times\boldsymbol{q})\cdot\boldsymbol{k}\,,\\
    && ig(q^2) \bar{u}(p_1)\gamma^\mu u(p_2) \frac{-ig_{\mu\nu}}{q^2+m_\rho^2} i\bar{u}(p_3)\sqrt{2}g_{\rho NN}\gamma^\nu u(p_4) \nonumber\\
    &&\approx -ig(-\boldsymbol{q}^2)4\sqrt{2}m_{P_c}m_Ng_{\rho NN} \frac{1}{\boldsymbol{q}^2+m_\rho^2} \nonumber \\
    &&\quad +i\sqrt{2} g(-\boldsymbol{q}^2)g_{\rho NN} \frac{1}{\boldsymbol{q}^2+m_\rho^2} \bigg{[} -4\boldsymbol{k}^2 + 2i((\boldsymbol{\sigma}_1+\boldsymbol{\sigma}_2)\times\boldsymbol{q})\cdot \boldsymbol{k}  \nonumber\\
    &&\quad + (\boldsymbol{\sigma}_1\times\boldsymbol{q})\cdot(\boldsymbol{\sigma}_2\times\boldsymbol{q}) \bigg{]}\,,\\
    && ig(q^2) \bar{u}(p_1)\gamma^\mu u(p_2) \frac{-ig_{\mu\nu}}{q^2+m_\rho^2} i\bar{u}(p_3)\frac{f_{\rho NN}}{\sqrt{2}m_N}\sigma^{\nu\lambda}(-iq_\lambda)u(p_4) \nonumber\\
    &&\approx  -\sqrt{2} \frac{m_{P_c}}{m_N} g(-\boldsymbol{q}^2)f_{\rho NN} \frac{1}{\boldsymbol{q}^2+m_\rho^2} \left[ 2(\boldsymbol{\sigma}_2\times \boldsymbol{k})\cdot\boldsymbol{q} + i\boldsymbol{q}^2 \right]   \nonumber\\
    &&\quad +\sqrt{2} g(-\boldsymbol{q}^2)f_{\rho NN} \frac{1}{\boldsymbol{q}^2+m_\rho^2} \nonumber\\
    &&\quad\times\left[ 2(\boldsymbol{\sigma}_2\times\boldsymbol{q})\cdot\boldsymbol{k} - i(\boldsymbol{\sigma}_1\times\boldsymbol{q})\cdot(\boldsymbol{\sigma}_2\times\boldsymbol{q}) \right]\,,\\
    &&ig(q^2) \bar{u}(p_1)u(p_2) \frac{i}{q^2-m_\sigma^2} ig_{\sigma NN} \bar{u}(p_3)u(p_4)   \nonumber\\
    &&\approx  i 4m_{P_c}m_N g(-\boldsymbol{q}^2) g_{\sigma NN} \frac{1}{\boldsymbol{q}^2+m_\sigma^2} \;.
\end{eqnarray}
Here, the momentum $\boldsymbol{k}$ and $\boldsymbol{q}$ are related to $\boldsymbol{p}_i$ by the following equation:
\begin{eqnarray}
\boldsymbol{p}_1+\boldsymbol{p}_2 &=& -(\boldsymbol{p}_3+\boldsymbol{p}_4)= 2\boldsymbol{k}\,,\\
\boldsymbol{p_2}-\boldsymbol{p_1} &=& \boldsymbol{p}_3-\boldsymbol{p}_4 = \boldsymbol{q}\,.
\end{eqnarray}}
After obtaining these scattering amplitudes, the effective potential in the  momentum space can be obtained by the Breit approximation~\cite{Berestetskii:1982qgu} using the equation 
\begin{equation}
  \mathcal{V}(\boldsymbol{q}) = -\frac{1}{2m_N2m_{P_c}}\mathcal{M}(\boldsymbol{q}) \;.
\end{equation}
Finally, the effective potential in the coordinate space can be obtained by the Fourier transformation,
\begin{equation}
\mathcal{V}(r) = \int \frac{d^3q}{(2\pi)^3} e^{i\boldsymbol{q}\cdot \boldsymbol{r}}\mathcal{V}(\boldsymbol{q})\mathcal{F}_M^2(\boldsymbol{q}^2,m_E^2)\;,
\end{equation}
where the monopole-type form factor $\mathcal{F}_M(\boldsymbol{q}^2,m_E^2)=(\Lambda^2-m_E^2)/(\Lambda^2+\boldsymbol{q}^2)$ is introduced to account for the inner structures of the hadrons and the off-shell nature of the exchanged light mesons, $m_E$ is the exchanged meson mass, and $\Lambda$ is the cutoff parameter. 
This kind of form factor is used to study the properties of deuteron~\cite{Machleidt:1987hj,Tornqvist:1993ng,Tornqvist:1993vu,Chen:2017jjn}, which indicates that a loosely bound state is a more probable candidate for the molecular state, with the cutoff parameter in the monopole-type form factor of approximately 1 GeV.
For the $P_c(4440)N$ and $P_c(4457)N$ systems, we can also obtain their effective interactions through the process described above.

\section{numerical results}

\begin{figure}[htbp]
  \centering
  \includegraphics[width=8.6cm]{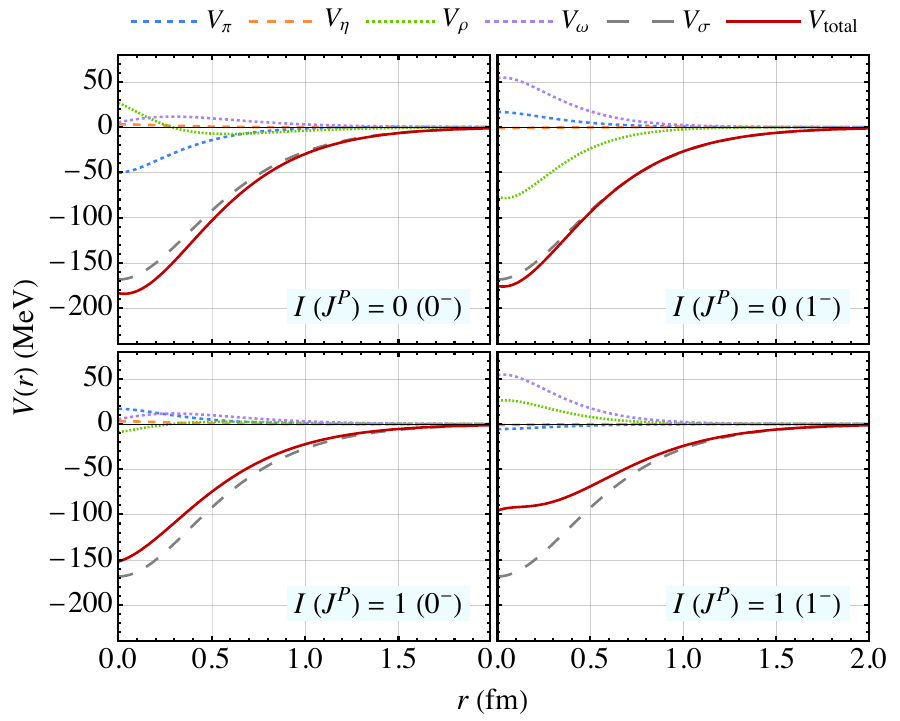}
  \caption{The effective interactions for the $P_c(4312)N$ system, where the cutoff parameter $\Lambda$ is taken to be the typical value of 1 GeV.}\label{fig:potential4312}
\end{figure}

With the method discussed in previous section, the potentials for the $P_cN$ systems are obtained.
For the $P_c(4312)N$ system, the corresponding $S$-wave effective potentials are shown in Fig.~\ref{fig:potential4312}. As illustrated in Fig.~\ref{fig:potential4312}, the $\pi$ and $\rho$ exchange potentials are a little more complex, contingent on both isospin and spin. The $\omega$ and $\eta$ exchange potential is dependent on the spin, and the $\eta$ exchange potential is small. The $\sigma$ exchange potential is independent of both isospin and spin, and provides the attractive potential \cite{Chen:2017vai}.
{The $\sigma$ meson exchange is observed to play a crucial role, providing significant attraction. In the OBE framework for nucleon-nucleon interactions~\cite{Machleidt:1987hj}, $\sigma$ meson exchange introduces medium-range attraction, effectively modeling the two-pion ($2\pi$) exchange process in effective field theory. This mechanism is a key component of the $NN$ interaction and is also instrumental in describing various hadron-hadron interactions~\cite{Chen:2017vai,Liu:2011xc}.}
In general, the $I=0$ $P_c(4312) N$ system is more attractive than the $I=1$ $P_c(4312) N$ system, which is similar to the effective potentials of the $\Sigma_c\bar{D}^{(*)}$ states, i.e., the $\Sigma_c\bar{D}^{(*)}$ states with the lower isospin are more tightly bound than the higher isospin states \cite{Chen:2019asm}.

\begin{figure}[htbp]
  \centering
  \includegraphics[width=8.4cm]{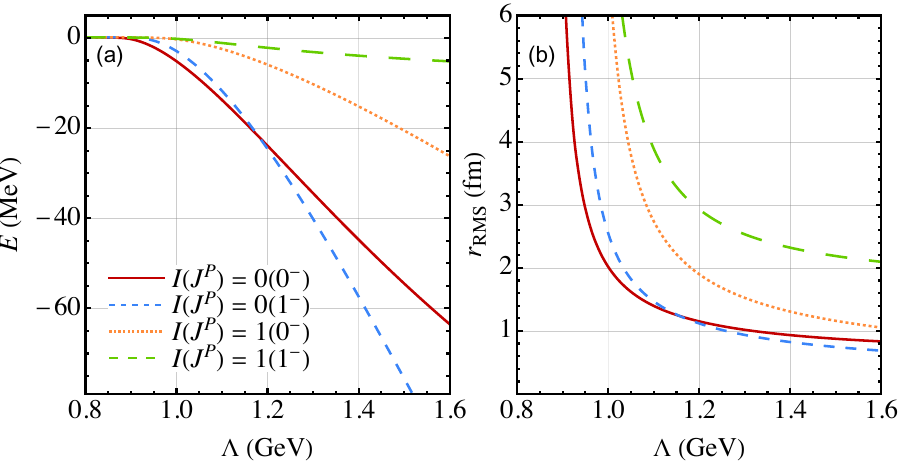}
  \caption{The bound state properties of the $P_c(4312)N$ system. (a) The binding energy under different cutoff parameter $\Lambda$. (b) The root-mean-square radius as the function of $\Lambda$.}\label{fig:BE4312}
\end{figure}

Furthermore, we discuss the bound state properties of the $P_c(4312)N$ system by solving the Schr\"{o}dinger equation. In Fig.~\ref{fig:BE4312}, we present the bound state properties of the $P_c(4312)N$ system, where the loosely bound state solution is searched by varying the cutoff $\Lambda$ in the range $0.8$-$1.6$ GeV. For the $I=0$ $P_c(4312)N$ system, the bound states emerge when the cutoff parameter is approximately equal to 1 GeV for both the $J=0$ and $J=1$ states. For the $I=1$ $P_c(4312)N$ system, the effective potentials are less attractive than those in the $I=0$ $P_c(4312)N$ system (see Fig.~\ref{fig:potential4312}), and our obtained numerical results indicate that the bound states can appear for the $I = 1$ $P_c(4312)N$ states with the relatively larger $\Lambda$ than the $I=0$ $P_c(4312)N$ states. For the $I(J^P)=1(1^-)$ $P_c(4312)N$ state, the shallow binding energy can obtain with the cutoff around 1.05 GeV, while the larger binding energy need a large $\Lambda$.
Based on studies of the deuteron~\cite{Machleidt:1987hj,Tornqvist:1993ng,Tornqvist:1993vu,Chen:2017jjn}, a loosely bound state is generally considered a more likely candidate for a molecular state, with the cutoff parameter in a monopole-type form factor typically around 1 GeV. Moreover, as the cutoff $\Lambda$ increases, the suppression of low-momentum contributions in the Fourier transformation diminishes, leading to a more attractive potential and a more tightly bound state for an attractive potential. From Fig.~\ref{fig:BE4312}, it is evident that the binding energy of the $I=0$ $P_cN$ system exhibits a more pronounced increase with increasing $\Lambda$ compared to the $I=1$ system. This indicates that the $I=0$ $P_cN$ system is more favorable for forming a bound state than the $I=1$ system.

{In our calculations, we employ the monopole-type form factor $\mathcal{F}_M(\boldsymbol{q}^2,m_E^2)=(\Lambda^2-m_E^2)/(\Lambda^2+\boldsymbol{q}^2)$ in the Fourier transformation, with a cutoff parameter $\Lambda$ set to 1 GeV, which is a reasonable choice for this type of form factor. However, the numerator term 
$(\Lambda^2-m_E^2)$ in this form factor significantly suppresses the vector meson exchange potential, especially for cutoff values below 1 GeV. To test the reliability of the present results, we also consider an alternative form factor $\mathcal{F}(\boldsymbol{q}^2)=\Lambda^2/(\Lambda^2+\boldsymbol{q}^2)$, which avoids the suppression of vector meson exchange contributions. For this alternative form factor, a reasonable choice for the cutoff parameter is $\Lambda=0.5$ GeV, as suggested in studies of hadronic molecular states~\cite{Chen:2017vai}.
Table~\ref{tab:be4312} summarizes the bound state properties of the $P_c(4312)N$ system using both types of form factors.}
\begin{table}
  \caption{\label{tab:be4312}
  The bound state properties of the $P_c(4312)N$ system under two different form factors. Case I represents the results obtained using the form factor $\mathcal{F}_M(\boldsymbol{q}^2,m_E^2)=(\Lambda^2-m_E^2)/(\Lambda^2+\boldsymbol{q}^2)$, while Case II corresponds to the results obtained with the form factor $\mathcal{F}(\boldsymbol{q}^2)=\Lambda^2/(\Lambda^2+\boldsymbol{q}^2)$.}
  \begin{ruledtabular}
    \begin{tabular}{cddd|ddd}
      & \multicolumn{3}{c|}{Case I} & \multicolumn{3}{c}{Case II} \\
      $I(J^P)$ & \multicolumn{1}{c}{$\Lambda$ (GeV)} & \multicolumn{1}{c}{$E$ (MeV)} & \multicolumn{1}{c|}{$r_{\mathrm{RMS}}$ (fm)} 
      & \multicolumn{1}{c}{$\Lambda$ (GeV)} & \multicolumn{1}{c}{$E$ (MeV)} & \multicolumn{1}{c}{$r_{\mathrm{RMS}}$ (fm)} \\
      \hline
               & 0.91 & -0.47 & 5.58   & 0.23 & -0.45 & 6.42  \\ 
      $0(0^-)$ & 1.02 & -6.67 & 1.83   & 0.38 & -6.46 & 2.17  \\
               & 1.16 & -19.75 & 1.23  & 0.54 & -19.66 & 1.41 \\
      \hline 
               & 0.94 & -0.33 & 6.56   & 0.23 & -0.43 & 6.55  \\
      $0(1^-)$ & 1.04 & -5.84 & 1.92   & 0.38 & -6.86 & 2.10  \\
               & 1.16 & -19.08 & 1.23  & 0.51 & -19.28 & 1.40 \\
      \hline
               & 1.01 & -0.42 & 5.92   & 0.40 & -0.41 & 6.30  \\
      $1(0^-)$ & 1.20 & -5.91 & 1.90   & 0.66 & -6.28 & 1.97  \\
               & 1.48 & -19.42 & 1.19  & 0.96 & -19.39 & 1.24 \\
      \hline
      \multirow{2}{*}{$1(1^-)$} & 1.04 & -0.49 & 5.54   & 0.48 & -0.51 & 5.74  \\
               & 1.50 & -4.59 & 2.19   & 1.00 & -4.30 & 2.34  \\
    \end{tabular}
  \end{ruledtabular}
\end{table}
{It is observed that the bound state properties for the two types of form factors are qualitatively similar. For the $I=0$ and $1(0^-)$ $P_c(4312)N$ systems, using the form factor $\mathcal{F}_M(\boldsymbol{q}^2,m_E^2)=(\Lambda^2 - m_E^2)/(\Lambda^2 + \boldsymbol{q}^2)$, bound states emerge when $\Lambda \approx 1$ GeV. In contrast, for the form factor $\mathcal{F}(\boldsymbol{q}^2)=\Lambda^2/(\Lambda^2 + \boldsymbol{q}^2)$, bound states appear at a lower cutoff, $\Lambda \approx 0.4$–$0.5$ GeV. For the $1(1^-)$ system, both form factors indicate that achieving a deep bound state requires a significantly larger cutoff $\Lambda$. This suggests that the $1(1^-)$ $P_c(4312)N$ system is relatively difficult to form a bound state.}

In addition to the $P_c(4312)N$ system, the loosely bound states can also exist for the $P_c(4440)N$ and $P_c(4457)N$ systems. {In Table \ref{Tab:BE44404457}, we present the bound state properties of the $P_c(4440)N$ and $P_c(4457)N$ systems when the cutoff $\Lambda$ varies from $0.8$ to $1.2$ GeV. For the $I=0$ $P_c(4440)N$ system, the presence of very shallow bound states is observed when the cutoff is set to $\Lambda=1.0$ GeV, and the binding energy increases when the cutoff is adjusted to $\Lambda=1.2$ GeV, this indicate the potential existence of the bound states. For the $P_c(4457)N$ system, the $I=0$ bound states can appear when the cutoff $\Lambda\approx 1$ GeV, while the $I=0$ bound states remains shallow when $\Lambda$ take $1.2$ GeV.} Consequently, the loosely bound states are more likely appear in the $I=0$ $P_c(4440)N$ and $P_c(4457)N$ systems compared to the $I=1$ systems, which is similar to the $P_c(4312)N$ system. 
\begin{table}
  \caption{\label{Tab:BE44404457}
  The bound state properties of the $P_c(4440)N$ and $P_c(4457)N$ systems. The $\Lambda$, $E$ and $r_{\mathrm{RMS}}$ are in unit of GeV, MeV and fm, respectively.}
  \begin{ruledtabular}
    \begin{tabular}{cddd|cddd}
      \multicolumn{4}{c|}{$P_c(4440)N$} & \multicolumn{4}{c}{$P_c(4457)N$} \\
      $I(J^P)$ & \multicolumn{1}{c}{$\Lambda$} & \multicolumn{1}{c}{$E$} & \multicolumn{1}{c|}{$r_{\mathrm{RMS}}$} 
      & $I(J^P)$ & \multicolumn{1}{c}{$\Lambda$} & \multicolumn{1}{c}{$E$} & \multicolumn{1}{c}{$r_{\mathrm{RMS}}$} \\
      \hline
               & 0.8 & \times & \times &          & 0.8 & \times & \times  \\ 
      $0(0^-)$ & 1.0 & -0.09  & 9.80   & $0(1^-)$ & 1.0 & -7.56  & 1.76  \\
               & 1.2 & -18.83 & 1.15   &          & 1.2 & -36.54 & 0.99 \\
      \hline 
               & 0.8 & \times & \times &          & 0.8 & \times & \times  \\
      $0(1^-)$ & 1.0 & -0.63  & 4.87   & $0(2^-)$ & 1.0 & -4.67  & 2.14  \\
               & 1.2 & -24.85 & 1.03   &          & 1.2 & -25.49 & 1.14 \\
      \hline
               & 0.8 & \times & \times &          & 0.8 & \times & \times  \\
      $1(0^-)$ & 1.0 & \times & \times & $1(1^-)$ & 1.0 & -1.26  & 3.67  \\
               & 1.2 & \times & \times &          & 1.2 & -7.95  & 1.73 \\
      \hline
               & 0.8 & \times & \times &          & 0.8 & \times & \times  \\
      $1(1^-)$ & 1.0 & \times & \times & $1(2^-)$ & 1.0 & -0.81  & 4.46  \\
               & 1.2 & \times & \times &          & 1.2 & -3.68  & 2.39
    \end{tabular}
  \end{ruledtabular}
\end{table}

\section{summery}
In the past decades, numerous new hadronic states were observed by different experiments, some of these new hadrons have stimulated extensive study of the hadronic molecules, such as $P_c(4312)$, $P_c(4440)$, $P_c(4457)$, $T_{cc}(3875)$, and so on, which are closely related to the interactions between the conventional hadrons. There is no end to the exploration of the matter world, and the interactions between such molecular states and the conventional hadrons may result in a variety of novel exotic hadronic matters. 

In this work, we propose a novel and universal mechanism to discuss the interactions between the molecular states and the conventional hadrons, i.e., the hadronic molecule and conventional hadron interactions can occur through the interactions between the constituent hadrons and the conventional hadrons. As a representative example, we mainly discuss the $P_cN$ interactions, and the $P_c$ states as the $\Sigma_c\bar{D}^{(*)}$ molecules can interact with the nucleon via the $\Sigma_cN$ and $\bar{D}^{(*)}N$ interactions.  Furthermore, we predict a new form of exotic hadronic matters inspired by the $P_cN$ interactions, and our numerical results suggest that the isospin-$0$ $P_cN$ systems are more likely to form the loosely bound states, which can be regarded as the most promising hidden-charm molecular octaquark candidates. 

Certainly, the mechanism proposed in the present work is universally applicable, which provides a good start point and new insight for discussing the molecular state and conventional hadron interactions and identifying a variety of novel exotic hadronic matters composed of the molecular states and the conventional hadrons.

\begin{acknowledgments}
This work is supported by the National Natural Science Foundation of China under Grant Nos. 12335001, 12247155, 12247101, and 12405097, National Key Research and Development Program of China under Contract No. 2020YFA0406400, the ‘111 Center’ under Grant No. B20063, the Natural Science Foundation of Gansu Province (No. 22JR5RA389), the fundamental Research Funds for the Central Universities, and the project for top-notch innovative talents of Gansu province.
\end{acknowledgments}

\bibliography{PcNbib}

\end{document}